\documentclass[draftclsnofoot, onecolumn, 12pt]{IEEEtran} 
\IEEEoverridecommandlockouts
\usepackage{cite}
\usepackage{amsmath,amssymb,amsfonts,comment,lipsum}
\usepackage{algorithmic}
\usepackage{graphicx}
\usepackage{textcomp}
\usepackage{xcolor}
\usepackage{hyperref}
\usepackage{algorithm}
\usepackage[utf8]{inputenc}
\usepackage{amsthm}

\def\BibTeX{{\rm B\kern-.05em{\sc i\kern-.025em b}\kern-.08em
    T\kern-.1667em\lower.7ex\hbox{E}\kern-.125emX}}

\begin{document}

\title{Acoustic RIS for Massive Spatial Multiplexing: Unleashing Degrees of Freedom and Capacity in Underwater Communications
\\
}

\author{
    \IEEEauthorblockN{
        Longfei Zhao\textsuperscript{1}, Jingbo Tan\textsuperscript{1}, Jintao Wang\textsuperscript{1}, Ian F. Akyildiz\textsuperscript{2}, Zhi Sun\textsuperscript{1§}
    }
    \IEEEauthorblockA{
        \textsuperscript{1}\textit{Department of Electronic Engineering}, Tsinghua University, Beijing, China \\
        Emails: zhaolf23@mails.tsinghua.edu.cn, tanjingbo@tsinghua.edu.cn, wangjintao@tsinghua.edu.cn, zhisun@ieee.org
    }

    \IEEEauthorblockA{
        \textsuperscript{2}\textit{Center for Robotics and Wireless Communications in Challenging Environments, Faculty of Electrical and Computer Engineering}, \\
        University of Iceland IS102 Reykjavik, Iceland \\
        Email: ianaky@hi.is
    }
    \vspace{-11mm}
    \thanks{\textsuperscript{§}Corresponding author: Zhi Sun (zhisun@ieee.org).
    
    }
}

\maketitle

\vspace{-18mm}
\begin{center}
    \fbox{\parbox{\dimexpr\linewidth-2\fboxsep-2\fboxrule\relax}{\centering
    This work has been accepted for publication at IEEE INFOCOM 2026.}}
\end{center}

\begin{abstract}
Underwater acoustic (UWA) communications are critically important for high-speed data transmission in marine applications, but are fundamentally constrained by limited bandwidth, severe propagation loss, and sparse multipath propagation. 
Conventional underwater acoustic multiple-input multiple-output (MIMO) systems rely on spatial diversity, yet their performance is inherently restricted by finite array resolution, leading to angular ambiguity and insufficient spatial degrees of freedom (DoFs). 
In this paper, these limitations are addressed through the novel application of acoustic Reconfigurable Intelligent Surfaces (aRIS), which are employed to actively generate orthogonally distinguishable virtual paths, thereby substantially increasing spatial DoFs and channel capacity. 
First, an ocean-specific DoF–channel coupling model is formulated, establishing explicit conditions for spatial rank enhancement.
Subsequently, the optimal geometric locus denoted as the \textit{Light-Point}, is analytically derived, where the strategic placement of a single aRIS maximizes spatial DoFs by introducing two and three additional resolvable paths in deep-sea and shallow-sea environments, respectively. 
Furthermore, an active simultaneous transmitting and reflecting (ASTAR) aRIS architecture is proposed, featuring independent beam control, alongside an adaptive beam-tracking mechanism that integrates unmanned underwater vehicles (UUVs) and acoustic intensity gradient sensing to maintain optimal performance under dynamic channel conditions.
The efficacy of the proposed joint aRIS deployment and beamforming framework is rigorously validated through extensive simulations, demonstrating remarkable improvements in UWA channel capacity up to $265\%$ in shallow-sea and $170\%$ in deep-sea scenarios.
These results underscore the transformative potential of aRIS technology in overcoming the intrinsic limitations of underwater acoustic communications.    
\end{abstract}

\section{Introduction}
Underwater acoustic (UWA) communications play a pivotal role in enabling oceanographic exploration and environmental monitoring, driving an ever-increasing demand for robust high-speed data transmission solutions\cite{beijing,11278690}. 
Emerging applications such as autonomous underwater robotics, real-time distributed sensing, and high-resolution seabed mapping further underscore the critical need for enhanced UWA communication capacity\cite{li2025underwater,10969775}.
More bandwidth-hungry missions—ranging from compressed underwater video feeds and rich telemetry to coordinated multi-robot operations—are already pushing UWA links to data rates on the order of $\ge$10~kbps (often tens of kbps and beyond), with high-definition video requiring substantially higher throughput\cite{1283454}.

To address the growing throughput requirements of UWA networks, recent research has focused on multiple-input multiple-output (MIMO) technology\cite{zhaoyangMIMO,men2025efficientresourceallocationmultiuser}, which exploits spatial diversity and multiplexing to significantly improve spectral efficiency, thereby enabling high-rate parallel data transmission within the constrained acoustic bandwidth. However, the intrinsic sparsity of UWA channels, coupled with the pronounced curvature of  the acoustic wavefronts, often results in signals arriving at receiving arrays with nearly identical angles of incidence. 
Furthermore, practical limitations on array size and deployment costs restrict the achievable spatial resolution, impeding the ability of conventional arrays to resolve closely spaced multipath components. 
Consequently, these systems primarily rely on spatial diversity to enhance the signal-to-noise ratio (SNR), leaving the potential capacity gains from spatial multiplexing largely untapped \cite{lowrank,MIMOdiversity}.
\begin{figure*}[htbp]
\centerline{\includegraphics[width=0.98\textwidth]{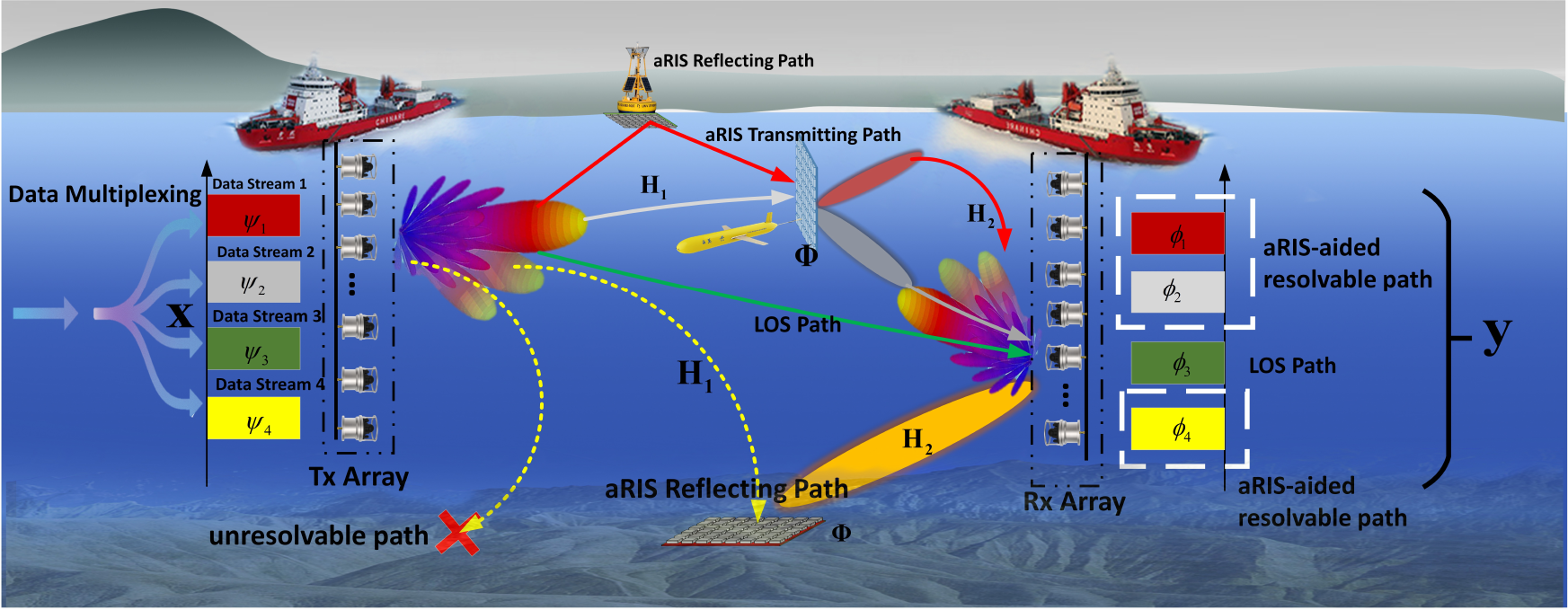}}
\vspace{-2mm} 
\caption{Increasing Underwater Channel capacity and Dof using Acoustic RIS.}
\label{fig}
\vspace{-7mm} 
\end{figure*}

The spatial multiplexing gain quantified as the spatial degrees of freedom (DoFs) determines the number of independent communication paths a MIMO channel can support. 
A higher DoF enables the concurrent transmission of multiple independent data streams, directly increasing channel capacity and throughput. 
However, the limited DoF in UWA MIMO systems severely constrains their spectral efficiency and overall performance\cite{ozgur2013spatial}.
As illustrated in Fig. 1, a typical high-speed UWA MIMO scenario involves multiple spatially multiplexed data streams. 
However, due to the restricted spatial resolution of practical arrays, the broad angular spread of acoustic propagation, and the absorption of high-angle emissions by the seabed (yellow dashed line, Fig. 1), the UWA channel often exhibits only a single dominant path typically the line-of-sight (LOS) component (green line, Fig. 1) \cite{zhang2025turbo}. 
This limitation drastically reduces the achievable DoF, as MIMO systems fundamentally depend on rich multipath scattering to establish resolvable transmission paths\cite{MIMOscatter,10623056}.

Given these inherent constraints, current UWA MIMO implementations have achieved only marginal improvements in data rates.
Recently, acoustic reconfigurable intelligent surfaces (aRIS) have emerged as a transformative solution, capable of actively manipulating acoustic wavefronts to enhance beamforming and signal propagation \cite{sun2022high,11161327}. 
Crucially, the dynamic reconfigurability of aRIS can also be leveraged to synthesize additional propagation paths \cite{10683678}, thereby augmenting the spatial diversity and multiplexing capabilities of UWA MIMO systems. 
However, despite this potential, no prior work has systematically explored the use of aRIS to increase the DoF of UWA channels.

In this paper, we propose for the first time in the literature the integration of aRIS into UWA networks to actively engineer multipath diversity, thereby increasing the rank and DoF of the MIMO channel matrix. As depicted in Fig. 1, aRIS can be seamlessly incorporated into existing infrastructure, such as surface buoys, seabed anchors, or unmanned underwater vehicles (UUVs)\cite{10978778,10143416,11153666}, enabling substantial throughput gains without extensive system redesign.

This paper makes the following key contributions to the field of underwater acoustic communications:

\begin{itemize}

\item \textbf{Optimal aRIS Deployment via Light-Point Analysis for Maximum Spatial Degrees of Freedom (DoFs):}
We present a comprehensive multipath-based system model for underwater acoustic MIMO communications assisted by active acoustic reconfigurable intelligent surfaces (aRIS). 
The model rigorously characterizes the enhancement of spatial DoF enabled by aRIS deployment. 
Building on this framework, we derive an optimal aRIS placement strategy that maximizes the achievable spatial DoF in underwater channels.
Central to this approach is the introduction of the \textit{Light-Point} concept defined as the geometric locus where distinguishable acoustic propagation paths intersect. 
We establish closed-form necessary and sufficient conditions for Light-Point existence and analytically prove that deploying a single aRIS at these points maximizes spatial DoF, yielding two additional resolvable paths in deep-sea environments and three in shallow-sea scenarios. 
Furthermore, we propose a two-stage deployment principle: first optimizing spatial rank through Light-Point identification, followed by minimizing total transmission loss to ensure both high DoF and robust link quality. 

\item \textbf{Independent Multi-DoF Beamforming via Underwater ASTAR-aRIS Architecture:}
To fully exploit the spatial DoF gains enabled by aRIS, we introduce an active simultaneous transmitting and reflecting (ASTAR) acoustic RIS architecture. 
Unlike conventional reflection-only designs, the ASTAR-aRIS independently controls transmission and reflection coefficients, enabling full 360° acoustic wave manipulation.
We develop an analytical electromagnetic-acoustic coupling model for the proposed structure, validated through high-fidelity COMSOL multiphysics simulations. 
Additionally, we present an independent multi-DoF beamforming strategy based on layered ray-tracing and second-order cone programming (SOCP) optimization. 
When integrated with the Light-Point deployment framework, this approach ensures effective multi-beam generation while maximizing spatial DoF.

\item \textbf{Dynamic Light-Point Tracking via Adaptive UUV-aRIS Integration:}
Addressing the inherent variability of underwater channels, we propose a novel adaptive beam-tracking mechanism that integrates UUVs with aRIS \cite{UUVRIS}.
By leveraging aRIS as an acoustic-intensity-aware sensing node, this system continuously monitors acoustic intensity gradients and dynamically repositions the aRIS to maintain optimal alignment with time-varying Light-Points. 
The proposed method ensures persistent high-capacity communication by compensating for environmental disturbances, such as water currents and temperature gradients, through real-time closed-loop adjustments.

\end{itemize}
\section{System Model}\label{Model}
We consider an underwater acoustic network, as illustrated in Fig.~1. The network nodes are equipped with multiple transducers to form UWA MIMO systems, aiming to achieve parallel high-speed data transmission. 
ARIS is deployed between the transmitter and receiver arrays, actively creating additional propagation paths (represented by red, orange, and gray curves in Fig.~1). 
$N_{t}$ projectors at the transmitter and $N_{r}$ hydrophones at the receiver are deployed in an array manner with normalized transducer spacings $\Delta_{t}$ and $\Delta_{r}$ respectively, measured in units of the carrier wavelength $\lambda_{c}$. 
Without loss of generality, it is assumed that the transmitter, receiver, and the aRIS located at the same vertical plane and the aRIS is perpendicular with this plane. 
The element number of aRIS is defined as $N_{a}\times N_{a}$ with $N_{a}$ representing the element number in the vertical or horizontal direction with spacing $\Delta_{a}$. 
In this way, the received signal $\mathbf{y}\in\mathcal{C}^{N_{r}\times 1}$ can be denoted as
\begin{equation}
\mathbf{y}=\left(\mathbf{H}+\mathbf{H}_{2}\mathbf{\Phi}\mathbf{H}_{1}\right)\mathbf{x}+\mathbf{n},
\end{equation}
where $\mathbf{x}\in\mathcal{C}^{N_{t}\times1}$ denotes the transmitted signals, $\mathbf{n}\in\mathcal{C}^{{N}_{r}\times 1}$ is the received noise, $\mathbf{H}\in\mathcal{C}^{{N_{r}}\times{N}_{t}}$, $\mathbf{H}_{1}\in\mathbf{C}^{N_{a}^{2}\times N_{t}}$, and $\mathbf{H}_{2}\in\mathcal{C}^{N_{r}\times N_{a}^{2}}$ are the channel between the receiver and the transmitter, the aRIS and the transmitter, and the receiver and the aRIS, respectively.
To accurately model and subsequently enhance the UWA channel capacity, the multipath channel model is utilized to describe the UWA MIMO channel\cite{tse2005fundamentals}. In specific, the channel $\mathbf{H},\mathbf{H}_{1},\mathbf{H}_{2}$ can be denoted as
\vspace{-2mm} 
\begin{subequations}\label{2}
\begin{equation}\label{2a}
\mathbf{H}=\sum\nolimits_{l=1}^{L}a_{l}\mathbf{e}_{r}\left(\phi_{l}\right)\mathbf{e}_{t}\left(\psi_{l}\right)^{H},
\end{equation}
\vspace{-4mm} 
\begin{equation}\label{2b}
\mathbf{H}_{1}=\sum\nolimits_{l=1}^{L_{1}}a_{1,l}\left(\mathbf{e}_{a}\left(\phi_{1,l}\right)\otimes\mathbf{1}_{N_{a}}\right)\mathbf{e}_{t}\left(\psi_{1,l}\right)^{H},
\end{equation}
\vspace{-2mm} 
\begin{equation}\label{2c}
\mathbf{H}_{2}=\sum\nolimits_{l=1}^{L_{2}}a_{2,l}\mathbf{e}_{r}\left(\phi_{2,l}\right)\left(\mathbf{e}_{a}\left(\psi_{2,l}\right)\otimes\mathbf{1}_{N_{a}}\right)^{H},
\end{equation}
\end{subequations}
where $L,L_{1},L_{2}$ denote the number of paths, $a_{l},a_{1,l},a_{2,l}$  denote the complex path gains, $\phi_{l},\phi_{1,l},\phi_{2,l}$ and $\psi_{l},\psi_{1,l},\psi_{2,l}$ are the path angle-of-arrivels (AoAs) and angle-of-departures (AoDs) in these channels, $\otimes$ represents the Kronecker product, $\mathbf{1}_{N_{a}}$ represents an all-ones vector of size $N_{a}\times 1$ and $\mathbf{e}_{t},\mathbf{e}_{r},\mathbf{e}_{a}$ are the steering vectors corresponding to the transmitter, receiver, and aRIS. The steering vector $\mathbf{e}_{t}$ has the form
\vspace{-2mm} 
\begin{equation}\label{3}
\begin{aligned}
\mathbf{e}_{t}&(\psi_{l})\!=\!\frac{1}{\sqrt{N_{t}}}\![1,\!e^{-j2\pi\Delta_{t}\psi_{l}},\!\cdots,\!e^{-j2\pi(N_{t}-1)\Delta_{t}\psi_{l}}]^{T},
\end{aligned}
\end{equation}
The steering vectors $\mathbf{e}_{r},\mathbf{e}_{a}$ have the same form with (\ref{3}) and can be obtained by substituting $N_{t}$ and $\Delta_{t}$ with $N_{r},N_{a}$ and $\Delta_{r},\Delta_{a}$. 
Specifically, the aRIS beamforming matrix $\mathbf{\Phi}\in\mathcal{C}^{N_{a}^{2}\times N_{a}^{2}}$represents the active control capability, independently adjusting both the amplitude and phase of incident acoustic waves. The beamforming matrix can be expressed as:
\vspace{-1mm} 
\begin{equation}\label{eq:RIS_beamforming}
    \mathbf{\Phi} = \text{diag}\{ \mu_1 e^{j\varphi_1},\, \mu_2 e^{j\varphi_2},\, \dots,\, \mu_{N_a^2} e^{j\varphi_{N_a^2}} \},
\end{equation}
where $\mu_n$ denotes the adjustable amplitude (including amplification), and $\varphi_n$ represents the tunable phase shift provided by the $n$-th element of the aRIS. This independent and simultaneous manipulation capability allows the aRIS to generate multiple distinct acoustic beams, as discussed in Sec.~\ref{BB}.

The DoF of MIMO channel, which is equal to the rank of the channel matrix, is an important metric in MIMO systems, since it describes how many streams can be simultaneously transmitted. Higher DoF usually means higher spatial multiplexing gain and thus higher channel capacity. In the multipath channel model as shown in (\ref{2}), the DoF, i.e., the rank of the channel matrix, has been proved to be equal to the number of resolvable paths. Therefore, for a UWA MIMO system without aRIS, its DoF is decided by the number of resolvable paths in $\mathbf{H}$, which can be denoted as
\begin{equation}\label{4}
\begin{aligned}
\mathrm{DoF}=\min\bigg(&\left|\left\{\psi_{l}|\cap_{k=1,k\neq{l}}^{L}\left|\psi_{l}-\psi_{k}\right|\geq\frac{1}{A_{t}}\right\}\right|,\\
&\left|\left\{\phi_{l}|\cap_{k=1,k\neq{l}}^{L}\left|\phi_{l}-\phi_{k}\right|\geq\frac{1}{A_{r}}\right\}\right|\bigg),
\end{aligned}
\end{equation}
where $\left|A\right|$ represents the cardinality of a set $A$, $\frac{1}{A_{t}},\frac{1}{A_{r}}$ are the angular resolution of the transmitting array and receiver array with $A_{t}=N_{t}\Delta_{t}$ and $A_{r}=N_{r}\Delta_{r}$ being aperture size.
As observed in (\ref{4}), the upper bound of the DoF is $L$. Unfortunately, because of limited scatters in the underwater environment and large absorption loss caused by sea surface and seabed, the number of paths $L$ in UWA MIMO channel is usually quite small, e.g., $L=1$ with single LOS path\cite{zhang2025turbo}. This leads to a low DoF and greatly limits the capacity that the UWA MIMO can achieve. In contrast, by introducing aRIS, the aRIS-aided UWA MIMO system may have the potential to increase DoF, since a new component of channel $\mathbf{H}_\mathrm{RIS}=\mathbf{H}_{2}\mathbf{\Phi}\mathbf{H}_{1}$ is constructed by aRIS. Therefore, the rank of $\mathbf{H}_{RIS}$ can be seen as the DoFs that introduced by aRIS which satisfies the following constraint as
\vspace{-1mm} 
\begin{equation}\label{5}
\mathrm{DoF}_\mathrm{RIS}\!=\mathrm{rank}\left(\mathbf{H}_\mathrm{RIS}\right)\!\leq\!\min\left(\mathrm{rank}\left(\mathbf{H}_{1}\right),\mathrm{rank}\left(\mathbf{H}_{2}\right)\right),
\end{equation}
where $\mathrm{rank}(\cdot)$ is the rank of a matrix. Considering $\mathbf{H}_\mathrm{RIS}=\mathbf{H}_{2}\mathbf{\Phi}\mathbf{H}_{1}$ and (\ref{5}), it can be concluded that two conditions are essential for achieving a larger $\mathrm{DoF}_\mathrm{RIS}$. First, the channels $\mathbf{H}_{1}$ and $\mathbf{H}_{2}$ have a higher rank, i.e., more resolvable paths. Second, the effective beamforming design enables the inequality in (\ref{5}) to achieve equality. Obviously, to realize the above two conditions, the deployment and beamforming design of aRIS should be well studied. However, relevant research remains scarce. Therefore, to fill this gap, we investigate the DoF-increasing ability of aRIS, and propose a joint deployment and beamforming scheme to maximize $\mathrm{DoF}_\mathrm{RIS}$ and improve channel capacity in the remaining sections.

\section{Optimal aRIS Deployment for Maximizing Underwater Acoustic Spatial Degrees of Freedom}\label{guideline}
Achieving the theoretical upper bound of spatial DoF in underwater UWA MIMO channels necessitates a meticulously optimized deployment and beamforming strategy for aRIS. The spatial DoF enhancement critically depends on two key factors: (1) the strategic placement of the aRIS to generate additional resolvable propagation paths, and (2) the precise design of beamforming coefficients to ensure these paths remain distinguishable at the receiver.
In this section, we analyze the deployment strategy for aRIS to maximize the upper bound of the spatial DoF $\mathrm{DoF}_\mathrm{RIS}$. 
Note that the proposed deployment principle is derived to optimize the DoF at the system-architecture level. Nevertheless, an important direction for future research is to study the impact of real-world constraints, such as power supply, deployment cost, and multi-user scheduling.

\vspace{-1mm}
\subsection{The aRIS Deployment Principle}
\label{sec:rank_enlargement}
From (\ref{5}), it can be known that the upper bound of $\mathrm{DoF}_\mathrm{RIS}$ is decided by $\mathrm{rank}(\mathbf{H}_{1})$ and $\mathrm{rank}(\mathbf{H}_{2})$, that is, $\sup \mathrm{DoF}_{RIS}=\min(\mathrm{rank}(\mathbf{H}_{1}),\mathrm{rank}(\mathbf{H}_{2}))$. Since the rank of $\mathbf{H}_{1}$ or $\mathbf{H}_{2}$ is equal to the number of resolvable propagation paths between the transmitter and the aRIS or between the aRIS and the receiver, the deployment position of aRIS that determines the AoDs and AoAs of the propagation paths is quite critical to the upper bound of DoF. Hence, in this subsection, we provide an aRIS deployment principle to enlarge the upper bound of DoF.

In specific, similar to (\ref{4}), the rank of the channel $\mathbf{H}_{1}$ and $\mathbf{H}_{2}$ can be denoted as\cite{tse2005fundamentals}
\vspace{-1mm}
\begin{subequations}\label{7}
\begin{equation}\label{7a}
\begin{aligned}
\mathrm{rank}\left(\mathbf{H}_{1}\right)\!=\!\min\bigg(&\left|\left\{\psi_{1,l}|\cap_{k=1,k\neq{l}}^{L_{1}}\left|\psi_{1,l}-\psi_{1,k}\right|\geq\frac{1}{A_{t}}\right\}\right|,\\
&\left|\left\{\phi_{1,l}|\cap_{k=1,k\neq{l}}^{L_{1}}\left|\phi_{1,l}-\phi_{1,k}\right|\geq\frac{1}{A_{a}}\right\}\right|\bigg),
\end{aligned}
\end{equation}
\begin{equation}\label{7b}
\begin{aligned}
\mathrm{rank}\left(\mathbf{H}_{2}\right)\!=\!\min\bigg(&\left|\left\{\psi_{2,l}|\cap_{k=1,k\neq{l}}^{L_{2}}\left|\psi_{2,l}-\psi_{2,k}\right|\geq\frac{1}{A_{a}}\right\}\right|,\\
&\left|\left\{\phi_{2,l}|\cap_{k=1,k\neq{l}}^{L_{2}}\left|\phi_{2,l}-\phi_{2,k}\right|\geq\frac{1}{A_{r}}\right\}\right|\bigg),
\end{aligned}
\end{equation}
\end{subequations}
where $A_{a}=N_{a}\Delta_{a}$. According to (\ref{7}), when an aRIS is deployed at an arbitrary candidate point $\mathbf{p}\triangleq(r,z)$ in the sea volume with fixed transmitter position and receiver position, the rank of $\mathbf{H}_{1}$ and $\mathbf{H}_{2}$ become related to the candidate point $\mathbf{p}$.
This is because the aRIS position determines the propagation paths, i.e., the sound ray, between the transmitter and the aRIS or between the aRIS and the receiver, and thus decides the number of resolvable paths. Therefore, based on (\ref{5}), the upper bound of DoF that increased by aRIS is also related to the aRIS deployment position $\mathbf{p}$, which can be represented as 
\vspace{-2mm}
\begin{equation}\label{8}
    \sup\mathrm{DoF}_\mathrm{RIS}(\mathbf{p}) \triangleq \min\!\bigl\{R_{1}(\mathbf{p}),\,R_{2}(\mathbf{p})\bigr\},
\end{equation}
where $R_{1}(\mathbf{p}),R_{2}(\mathbf{p})$ denote the rank of $\mathbf{H}_{1}$ and $\mathbf{H}_{2}$ when the aRIS is deployed at $\mathbf{p}$.
Intuitively, $R_{1}(\mathbf{p})$ counts how many angularly resolvable rays can depart from the transmitter and hit the candidate point~$\mathbf{p}$, whereas $R_{2}(\mathbf{p})$ counts how many such rays can leave~$\mathbf{p}$ and still reach the receiver with resolvable angles.
Scanning $\mathbf{p}$ over the region of interest produces a scalar field
$\sup\mathrm{DoF}_\mathrm{RIS}(\mathbf{p})$ that we refer to as the \emph{DoF map}.
Formally,
\begin{equation}
    \mathcal{S}_{\mathrm{DoF}}
    \;=\;
    \bigl\{\mathbf{p}\;|\;\sup\mathrm{DoF}_\mathrm{RIS}(\mathbf{p})\ge 1\bigr\},
\end{equation}
and points with identical values can be collected into iso‑DoF contours by
\begin{subequations}\label{10}
\begin{equation}
    \mathcal{S}_{\mathrm{DoF}}=\bigcup\nolimits_{d=1}^{d_{\max}}
        \mathcal{S}_{\mathrm{DoF}}^{(d)},
        \end{equation}
\begin{equation}
    \mathcal{S}_{\mathrm{DoF}}^{(d)}
    =\bigl\{\mathbf{p}\in\mathcal{S}_{\mathrm{DoF}}
      \,\bigl|\,\sup\mathrm{DoF}_\mathrm{RIS}(\mathbf{p})=d\bigr\},
\end{equation}
\end{subequations}
where $d_\mathrm{max}=\max_{\mathbf{p}}{\sup\mathrm{DoF}_\mathrm{RIS}(\mathbf{p})}$. In (\ref{10}), the set $\mathcal{S}_{\mathrm{DoF}}^{(d_{\max})}$ contains all the positions where an aRIS can inject the \emph{highest} number of potential parallel virtual channels, i.e., the largest upper bound of $\mathrm{DoF}_\mathrm{RIS}$. Therefore, it can be concluded that the principle to deploy aRIS is to make the aRIS located at one of the points within $\mathcal{S}_{\mathrm{DoF}}^{(d_{\max})}$. To highlight these points, we refer to points $\mathbf{p}\in\mathcal{S}_{\mathrm{DoF}}^{(d_{\max})}$ as the \emph{Light-Points}.

Further, DoF alone does not guarantee link quality, because propagation loss varies according to the position of the aRIS. The optimal deployment position of the aRIS should be the \emph{Light-Point} $\mathbf{p}_{\star}$ that minimizes the sum of transmission loss of paths built by the aRIS in $\mathbf{H}_\mathrm{RIS}$. Specifically, for any candidate point $\mathbf{p}$, let $\mathcal{L}_{1}(\mathbf{p})=\{\,a_{1,l}(\mathbf{p})\}_{l=1}^{R_{1}(\mathbf{p})}$ and  $\mathcal{L}_{2}(\mathbf{p})=\{\,a_{2,l}(\mathbf{p})\}_{l=1}^{R_{2}(\mathbf{p})}$ 
denote the complex path gains in \eqref{2b}–\eqref{2c}. Then, the one‑way transmission loss of each path in $\mathbf{H}_{1}$ and $\mathbf{H}_{2}$ can be obtained as 
\vspace{-1mm}
\begin{equation}\label{11}
TL_{i,l}(\mathbf{p})=20\log_{10}\left|a_{i,l}(\mathbf{p})\right|,
\end{equation}
where $i\in\{1,2\}$. Notice that the aRIS can integrate one path each in $\mathbf{H}_{1}$ and $\mathbf{H}_{2}$ into a single path of $\mathbf{H}_\mathrm{RIS}$ through beamforming. Therefore, 
To evaluate the transmission loss of the aRIS placement, we select \emph{any} $d\!=\!\sup\mathrm{DoF}_{\mathrm{RIS}}(\mathbf{p})$ mutually resolvable paths from each side, sum their losses, and minimize the summation to serve as a metric\cite{ainslie2010principles}, i.e.
\begin{equation}\label{12}
   TL(\mathbf{p})\triangleq
   \!\!\!\!\!\min_{\substack{%
        S\subseteq\mathcal{L}_{1}(\mathbf{p}),\;
        T\subseteq\mathcal{L}_{2}(\mathbf{p})\\
        |S|=|T|=d
      }}
   \Bigl(
        \sum_{a_{1,l}\in S}\!TL_{1,l}(\mathbf{p})
        +\!
        \sum_{a_{2,l}\in T}\!TL_{2,l}(\mathbf{p})
   \Bigr),
\end{equation}
so that all possible combinations of the $R_{1}(\mathbf{p})$ and $R_{2}(\mathbf{p})$ candidate rays are considered while retaining the $d$ rays required to realize the upper‑bound DoF at $\mathbf{p}$. Therefore, minimizing $TL(\mathbf{p})$ over $\mathcal{S}^{(d_{\max})}_{\mathrm{DoF}}$ yields the optimal \emph{Light-Point} $\mathbf{p}_{\star}$, which satisfies
\vspace{-2mm}
\begin{equation} 
    \mathbf{p}_{\star}
    \;=\;
    \arg\!\min_{\mathbf{p}\in\mathcal{S}_{\mathrm{DoF}}^{(d_{\max})}}
        TL(\mathbf{p}).
    \label{eq:light_point_def}
\end{equation}
The above two-stage procedure—first using the \emph{DoF map} to identify the most valuable region that can generate the maximum number of resolvable paths (i.e., the candidate set of Light-Points), and then performing a local search within this region to minimize the transmission-loss metric $TL(\mathbf{p})$—yields a physically transparent and computationally lightweight principle for aRIS placement, thereby avoiding a large-scale nonconvex mixed optimization over both $\mathbf{p}$ and $\boldsymbol{\Phi}$.

%

\subsection{Light-Point Model for aRIS Deployment}
Having established the two-stage deployment principle, the next critical step is to validate the existence of Light-Points and quantify the upper bound of DoF achievable when an aRIS is deployed at these Light Points. Given that acoustic propagation channels can be effectively modeled using sound rays, we demonstrate the existence of Light-Points and derive the DoF upper bound through rigorous ray analysis in both deep-sea and shallow-sea scenarios.

First, we focus on a mid- to long-range deep-sea scenario with transmission range \(R=\mathcal{O}(50\,\mathrm{km})\), where the large propagation span and the upward-refracting sound–speed profile (SSP) bend eigenrays into arch shapes. The following \textbf{Theorem 1} proves that in typical deep-sea environments, acoustic rays emitted from the transmitter—including boundary-reflected rays and symmetric direct rays refracted near the sound channel axis—naturally intersect at a common location, referred to as a Light-Point $\mathbf{p}$. At this Light-Point, an effective spatial DoF of 2 can be realized by exploiting these propagation paths. Considering the significant reflection loss at large incidence angles\cite{hamilton1976sound}, we restrict our subsequent analysis to rays undergoing only a single reflection.



\begin{figure}[htbp]
\centerline{\includegraphics[width=0.98\textwidth]{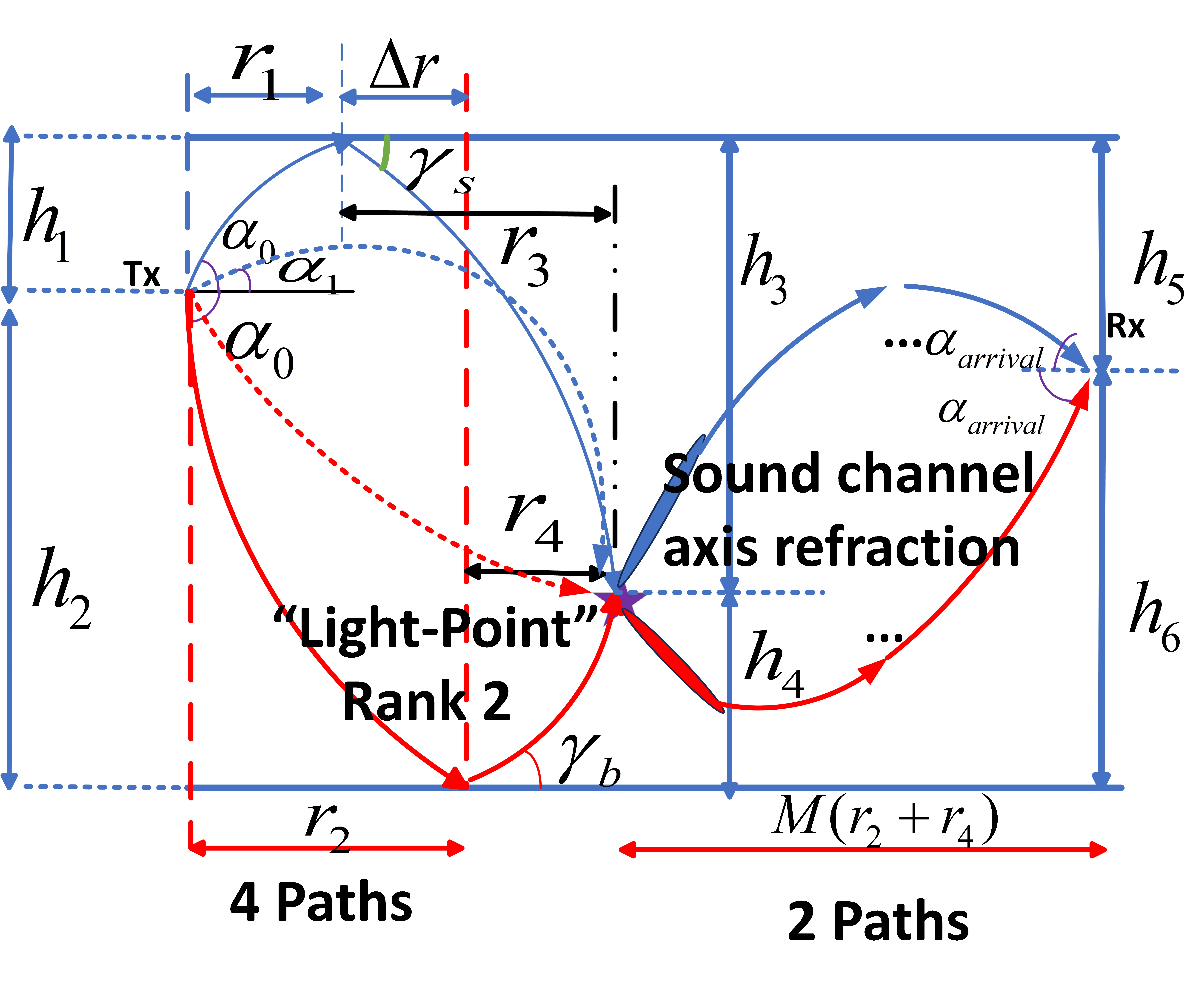}}
\vspace{-4mm}
\caption{The aRIS deployment in the typical Deep-Sea Scenario.}
\label{sh}
\vspace{-6mm}
\end{figure}

\noindent\textbf{Theorem 1.}
\textit{In typical deep-sea environments, four distinct acoustic rays comprising sea-surface-reflected, seabed-reflected, and two symmetric LOS paths converge at a common Light-Point $\mathbf{p}$ between the transmitter and the aRIS. While, due to the substantial propagation distances\cite{weizhe2025underwater}, only two symmetric direct rays (refracted near the sound channel axis without boundary reflections) exist between the aRIS and the receiver, resulting in two distinguishable paths. If the aRIS is positioned near the receiver, the condition becomes symmetric. For these Light-Points, the supremum achievable DoF $\sup\mathrm{DoF}_\mathrm{RIS}=2$.}

\noindent\textit{Proof.}
  Consider a typical deep-sea SSP (e.g., the Munk profile):
$c(z)=c_s\left[1+\epsilon(\eta+e^{-\eta}-1)\right]$, where $\eta=\frac{z-z_0}{z_l}$, $\epsilon\approx 7\times10^{-3}$, $z_0$ is the SOFAR axis depth, and $z_l$ is the scale depth. As shown in Fig.~2, due to the limited angular resolution ($1/A_t\approx20^{\circ}$) of practical acoustic arrays\cite{qiao2017mimo}, emitted acoustic rays can be categorized into two groups based on the critical grazing angle $\alpha_{\text{crit}}=\arccos({c_0}/{c_s})$, where $c_0$ and $c_s$ represent the sound speeds at the source depth and sea surface, respectively. These are: (1) boundary-reflected rays (angles $>\alpha_{\text{crit}}$) and (2) direct LOS rays (angles $<\alpha_{\text{crit}}$).

First, we verify the existence of Light-Points formed by boundary-reflected rays. The horizontal propagation distance difference for a ray emitted with grazing angle $\alpha_0$ before and after boundary reflection is:
\begin{equation}
    \Delta r=\cos\alpha_0\int_{0}^{h_1+h_2}\frac{dz}{\sqrt{n^2(z)-\cos^2\alpha_0}},
\end{equation}
where $n(z)=c_0/c(z)$ and $h_1,h_2$ are relevant depths. Similarly, for boundary-reflected rays at upper ($\gamma_s$) and lower ($\gamma_b$) grazing angles, the horizontal distance difference is:
\begin{equation}
    r_3 - r_4 \approx \Delta r,
\end{equation}
assuming typical deep-sea conditions ($|c_s/c_b-1|\ll1$)\cite{SSP}. This confirms stable intersections at Light-Points.

Next, consider direct LOS rays emitted at small symmetric grazing angles ($\alpha_1$, $-\alpha_1$). For any fixed Light-Point $\mathbf{p}$, a suitable $\alpha_1$ always exists satisfying:
\begin{equation}
    r_{\text{LOS}}(\alpha_1)=\cos\alpha_1\int_{h_1}^{h_{\mathbf{p}_\star}}
    \frac{dz}{\sqrt{n^2(z)-\cos^2\alpha_1}},
\end{equation}
where $h_{\mathbf{p}_\star}$ is the Light-Point depth. Due to symmetry of SSP around the SOFAR axis, we have:
\begin{equation}
    r_{\text{LOS}}(-\alpha_1)=r_{\text{LOS}}(\alpha_1).
\end{equation}
confirming that direct LOS rays intersect precisely at the same Light-Point $\mathbf{p}$. Thus, all four paths intersect at $\mathbf{p}$, yielding $\operatorname{rank}\mathbf{H}_{1}=4$.
Finally, from aRIS to receiver, due to significant spans, rays emitted at large angles experience multiple boundary reflections and substantial losses\cite{jia2025message}, becoming negligible. Effective acoustic paths are thus primarily symmetric refracted rays around the SOFAR channel axis (no boundary reflections). For these rays, the total horizontal distance after $M$ cycles is: $r_{\text{total}} = M(r_2 + r_4)$. Due to practical angular resolution ($1/A_r \approx 20^\circ$)\cite{qiao2017mimo}, only two symmetric refracted rays are distinguishable, giving $\mathrm{rank}\,\mathbf{H}_2=2$. By acoustic reciprocity\cite{fokkema2013seismic}, this result symmetrically applies when aRIS is near the receiver. Thus, the achievable DoF is determined as: $ \sup\mathrm{DoF}_\mathrm{RIS} = \min(\operatorname{rank}\mathbf{H}_1,\operatorname{rank}\mathbf{H}_2)=2$,
confirming $\sup\mathrm{DoF}_\mathrm{RIS}=2$ is realizable.\hfill$\blacksquare$

\begin{figure}[htbp]
\centerline{\includegraphics[width=0.98\textwidth]{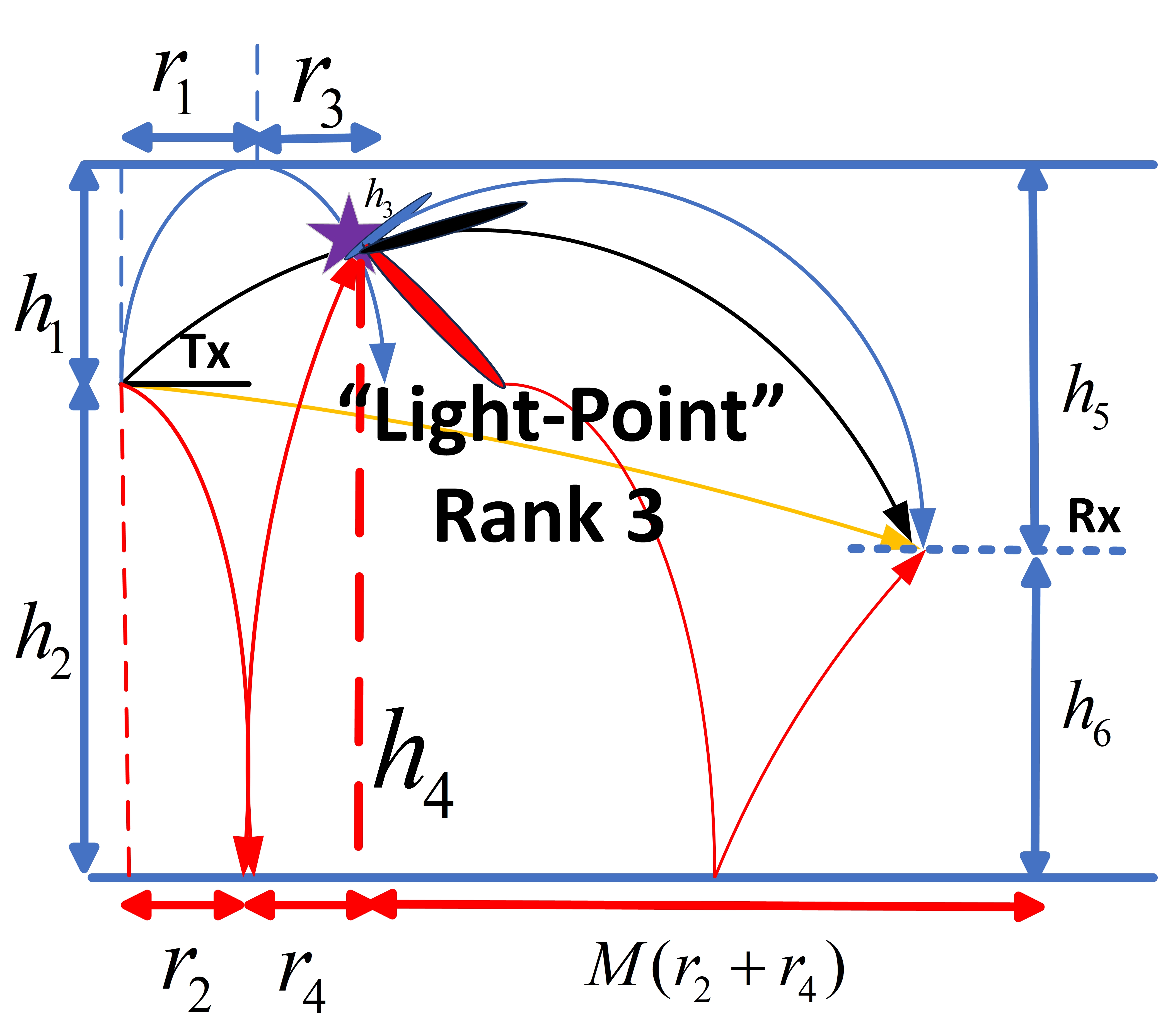}}
\vspace{-3mm} 
\caption{The aRIS deployment in the typical Shallow-Sea Scenario.}
\vspace{-6mm} 
\label{fig}
\end{figure}

Second, in a shallow‑sea waveguide the the transmitter–receiver range is moderate (\(R=\mathcal{O}(5\,\mathrm{km})\)) and the depth of the sea \(H=\mathcal{O}(100\,\mathrm{m})\).
Because the sea surface and seabed lie close to each other, an acoustic ray emitted from the transmitter can (i) arrive at the aRIS in LOS path, (ii) reach the aRIS after a single surface reflection, or (iii) after a single bottom reflection. These three rays inevitably cross in a compact region whose depth is between the two boundaries. \textbf{Theorem 2} illustrates such an intersection, as shown in Fig.3. 

\noindent\textbf{Theorem 2.}
\textit{In typical shallow-water environments, three dominant acoustic ray paths —the sea-surface-reflected path, the seabed-reflected path, and the LOS path— naturally converge at a focal point, termed a Light-Point $\mathbf{p}$, between either the transmitter and the aRIS or the aRIS and the receiver. At these Light-Points, the upper bound on the spatial DoF of the aRIS, denoted as $\sup\mathrm{DoF}_{RIS}$, can be achieved with a maximum value of 3.}

\noindent\textit{Proof.}  Using sound ray analysis method similar to \textbf{Theorem 1} and based on the geometry of a shallow-sea waveguide\cite{zhou2014ofdm}, we can prove the Light-Points described in \textbf{Theorem 2} exists, as shown in Fig.~3. It should be emphasized that unlike the deep‑sea scenario\cite{10.1121/1.4747617}, the short-range shallow‑water geometry always furnishes a LOS path between the transmitter and the aRIS or the aRIS and the receiver, which makes $\mathrm{rank}(\mathbf{H}_{1})=3$ and $\mathrm{rank}(\mathbf{H}_{2})=3$ can be realized simultaneously. In this way, when the aRIS is deployed at Light-Points, $\sup\mathrm{DoF}_\mathrm{RIS}=3$ naturally holds.\hfill$\blacksquare$

\section{Independent Multi-DoF Beamforming based on ASTAR aRIS Design
}\label{STARaRISDesign}
To fulfill the upper bound of DoF provided by aRIS, i.e., $\sup\mathrm{DoF}_\mathrm{RIS}=2$ in deep-sea scenario and $\sup\mathrm{DoF}_\mathrm{RIS}=3$ in shallow-sea scenario, the beamforming of aRIS should be carefully designed to integrate paths in $\mathbf{H}_{1}$ and $\mathbf{H}_{2}$ and construct resolvable paths in $\mathbf{H}_\mathrm{RIS}$. However, existing reflecting-only aRIS inherently constrains acoustic energy manipulation to only one side of the aRIS array, making it challenging to achieve full $360$-degree beamforming~\cite{INFOCOM2025,luo2025underwater}, especially at Light-Points $\mathbf{p}$ identified in Section III, which restricts the number of paths can be constructed in $\mathbf{H}_\mathrm{RIS}$ and makes $\sup\mathrm{DoF}_\mathrm{RIS}$ cannot be achieved. To solve the above challenge, we propose an ASTAR aRIS hardware structure that enables simultaneous transmission and reflection of acoustic waves. 

\subsection{ASTAR aRIS Hardware Design and Validation}\label{AA}
As illustrated in Fig. 4, the proposed hardware structure consists of an array of active piezoelectric elements integrated with reflection-type amplifiers and phase shift circuits. Each element independently splits the incoming acoustic wave into transmitted and reflected parts using a hybrid coupler with ports {1, 2, 3, 4}\cite{tsai2011miniaturized}. Specifically, the acoustic wave incident from port 1 is simultaneously output through ports 2 and 3, subsequently passing through two independent phase-shift circuits and reflection-type amplifiers with adjustable gain $G_{\mathrm{T},n}$ and $G_{\mathrm{R},n}$, respectively. Thus, the reflected ($\widetilde S_n$) and transmitted ($\widetilde T_n$) coefficient at the output ports (ports 1 and 4, respectively) of the hybrid coupler for the $n$-th aRIS element can be independently expressed as\cite{STARRIS}
\begin{equation}\label{23}
\widetilde{S}_n \!=\! j\frac{\widetilde{G_T} - \widetilde{G_R}}{2}\!=\!|R_n|e^{j\varpi_n},\quad\!\!\!\!\!\!\!
\widetilde{T}_n \!= \!j\frac{\widetilde{G_T} + \widetilde{G_R}}{2}=|T_n|e^{j\varphi_n}.
\end{equation}
By independently tuning gains and shifts of the complex ($\widetilde{G_T}$, $\widetilde{G_R}$) provided by each reflection-type amplifier, the amplitude and phase of transmitted and reflected acoustic waves can be individually controlled.
\begin{figure}[htbp]
\centerline{\includegraphics[width=0.98\textwidth]{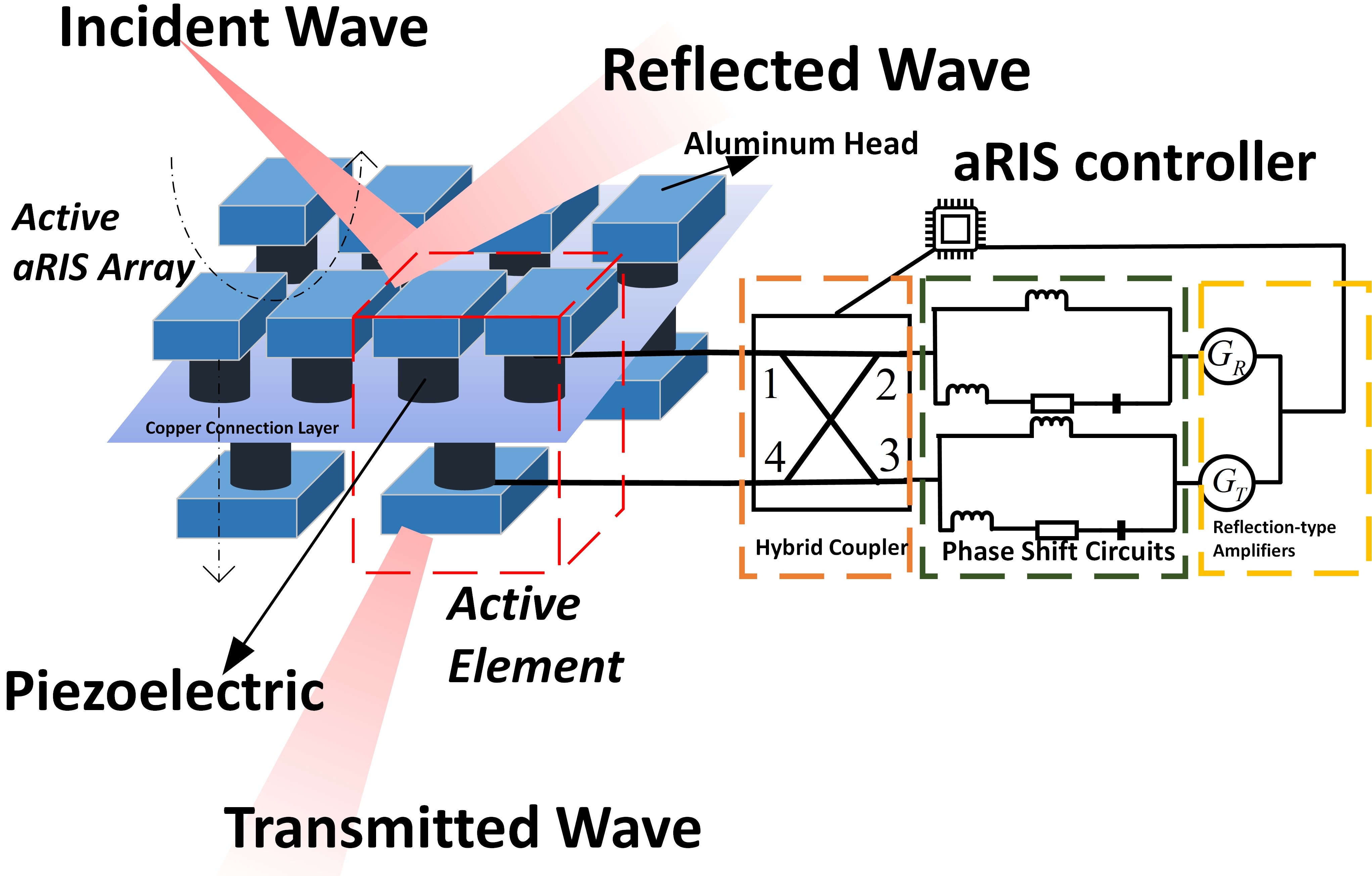}}
\vspace{-3mm} 
\caption{Hardware structure of the underwater ASTAR aRIS.}
\label{fig}
\vspace{-7mm}
\end{figure}
To verify the feasibility of the proposed ASTAR aRIS hardware structure, numerical simulations are performed in COMSOL under free-water  field condition~\cite{multiphysics1998introduction}. A coupled acoustic-electric model of piezoelectric materials is adopted, where a four-element aRIS array is constructed using PZT-4 piezoelectric rings operating at $9$ kHz. Each element, arranged with half-wavelength spacing, is connected to an independent phase-shift and amplification circuit~\cite{sherman2007transducers}. The incident acoustic plane wave has an amplitude of $100$ Pa and an angle of $45^\circ$. As shown in Fig.~5, when the aRIS is inactive, the acoustic pressure field remains unchanged. Upon activation, the ASTAR aRIS simultaneously generates a reflected beam at $80^\circ$ (on the opposite side of the incident wave) and a transmitted beam at $30^\circ$ (on the same side). Clear interference fringes appear in both directions, forming distinct beam patterns resulting from constructive and destructive interference, which confirms the directional control and effectiveness of our ASTAR aRIS design.
\begin{figure}[htbp]
\centerline{\includegraphics[width=0.98\textwidth]{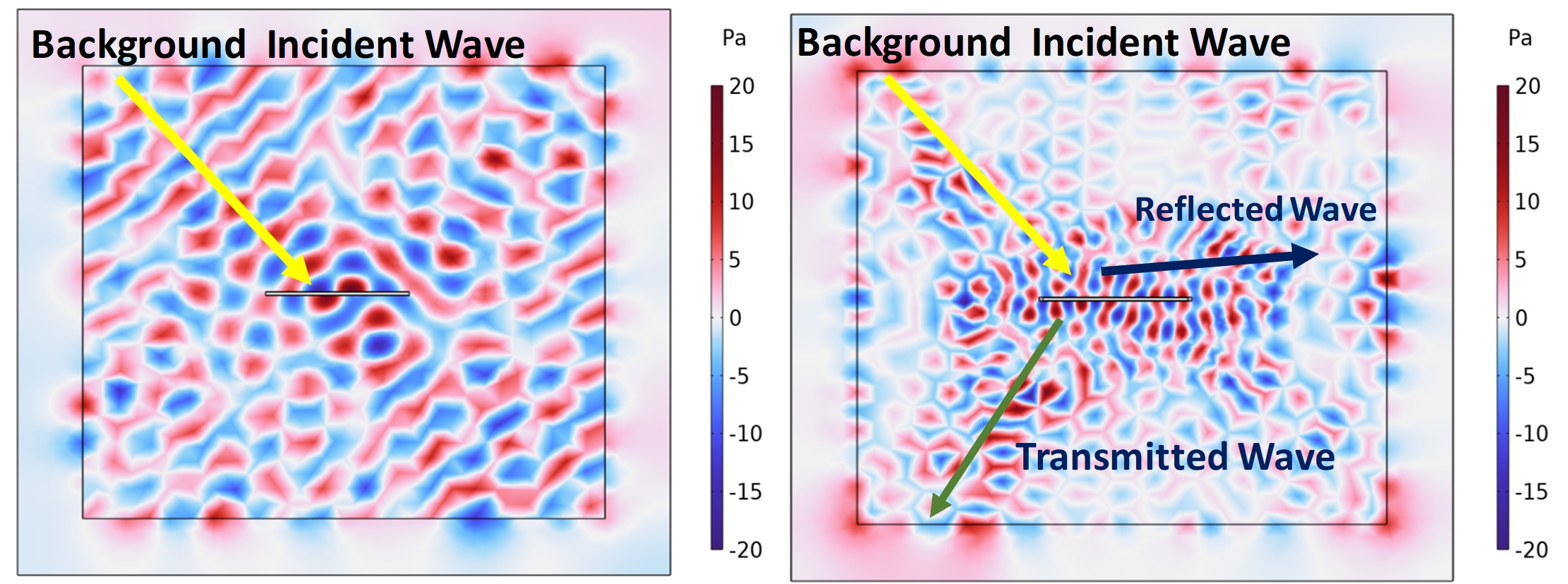}}
\vspace{-2mm} 
\caption{Simulated pressure field of the designed ASTAR aRIS in COMSOL.}
\vspace{-5mm} 
\label{fig}
\end{figure}
\vspace{-1mm}

\subsection{Independent Multi-DoF Beamforming}\label{BB}
Based on the transmission ability of the designed ASTAR aRIS, when it is deployed at the Light-Points as shown in Fig. 2 or Fig.3, the DoF provided by aRIS could reach its upper bound by efficient beamforming design. In this subsection, we formulate an optimization problem to realize independent multi-DoFs beamforming which can fulfill the upper bound of DoF provided by aRIS. Obviously, the key of multi-DoFs beamforming is to collect signals in separated paths from transmitter and launch \emph{multiple angularly resolvable beams} toward the receiver. 
Without loss of generality, we suppose only the transmitting function of the ASTAR aRIS is utilized in multi-DoF beamforming, which is feasible when aRIS is deployed at the Light-Points as have illustrated in Fig. 2 and Fig. 3. On the other hand, because the transmitter, aRIS, and receiver lie in the same vertical $(r,z)$ plane and the aRIS is normal to that plane (Sec.~\ref{Model}), the beamforming coefficients provided by aRIS elements in the same row in the direction that is perpendicular to the $(r,z)$ plane should be the same. Therefore, the beamforming matrix $\mathbf{\Phi}$ satisfies $\mathbf{\Phi}=\mathrm{diag}\{\mathbf{T}\otimes\mathbf{1}_{N_{a}}\}$ where $\mathbf{T}\in\mathcal{C}^{N_{a}\times 1}$ is the beamforming vector in the columns of the aRIS which are parallel to  the $(r,z)$ plane. In this way, the beamforming matrix design problem is equivalent to design a beamforming vector $\mathbf{T}$, just as a linear aRIS.
Following the idea of first collecting the signals of the incident waves and then forming multiple resolvable beams toward the receiver, we propose to separate the beamforming coefficients in $\mathbf{T}$ as $\mathbf{T}=[T_{1},T_{2},\dots,T_{N_{a}}]^{T}$ into two pars by $T_{t}=T_{\mathrm{c},t}T_{\mathrm{g},t}$. $T_{\mathrm{c},t}$ is utilized to capture acoustic signals from the transmitter, while $T_{\mathrm{g},t}$ is optimized to generate independent multi-DoF beams.
In specific, assuming that the number of resolvable paths that the aRIS should generate is $\sup\mathrm{DoF}_\mathrm{RIS}=P$ and the AoAs of $P$ incident paths at the aRIS are $\phi_{1,p}$ with $p=1,2,\cdots,P$, the coefficients $T_{\mathrm{c},t}$ can be calculated by
\begin{equation}\label{29}
T_{\mathrm{c},t}=\Sigma_{p=1}^{P}T_{\mathrm{c},t}^{(p)},\quad T_{\mathrm{c},t}^{(p)}=e^{j2\pi\Delta_{a}\phi_{1,p}},
\end{equation}
where $T_{\mathrm{c},t}^{p}$ denotes the phase-shifts that required to align and capture the $p$-th path for the $t$-th aRIS element.
\begin{algorithm}[htbp]
\caption{Joint aRIS Deployment and Beamforming}
\label{alg:joint}
\begin{algorithmic}[1]
    \STATE \textbf{Input:} Tx position $(r_t,z_t)$, Rx position $(r_r,z_r)$,
           SSP $c(z)$ discretised into $M$ layers $\{c_m,h_m\}_{m=1}^{M}$, arrya size parameters $N_{t},N_{r},N_{a}$, aRIS beamforming optimization parameters $G_{i}^{req},\epsilon_{cross},G_{max}$
    \STATE \textbf{Output:} Optimal Light-Point $\mathbf p_\star$, aRIS coefficients $\mathbf{\Phi}$

    \STATE $\mathcal S_{\max}\leftarrow\varnothing,\; d_{\max}\leftarrow 0$
    \STATE \textbf{for} each grid point $\mathbf p=(r,z)$ allowed by $\mathrm{Env}$ \textbf{do}
    \STATE  \quad $L_{1},\{\phi_{1,l}\},\{\psi_{1,l}\}\leftarrow$ LRT$\left((r_{t},z_{t}),\mathbf{p},c(z)\right)$
    \STATE  \quad $L_{2},\{\phi_{2,l}\},\{\psi_{2,l}\}\leftarrow$ LRT$\left(\mathbf{p},(r_{r},z_{r}),c(z)\right)$
    \STATE \quad Compute $R_1(\mathbf p),R_2(\mathbf p)$ via \eqref{7}
    \STATE \quad $d(\mathbf p)\leftarrow\min\{R_1(\mathbf p),R_2(\mathbf p)\}$
    \STATE \quad \textbf{if} $d(\mathbf p)>d_{\max}$ \textbf{then}\quad $d_{\max}\leftarrow d(\mathbf p), \mathcal S_{\max}\leftarrow\{\mathbf p\}$
    \STATE \quad \textbf{else if} $d(\mathbf p)=d_{\max}$ \textbf{then}\quad $\mathcal S_{\max}\leftarrow\mathcal S_{\max}\cup\{\mathbf p\}$
    \STATE \quad \textbf{end if}
    \STATE  \textbf{end for}
    \STATE $\mathbf p_\star\leftarrow\arg\min_{\mathbf p\in\mathcal S_{\max}} TL(\mathbf p)$ via \eqref{eq:light_point_def}
    \STATE Obtain $\{\phi_{1,l}^{\star}\}$ and $\{\phi_{2,l}^{\star}\}$ using LRT similar to step 5 and 6 by replacing $\mathbf{p}$ with $\mathbf{p}_{\star}$
    \STATE Select $\{\psi_{1,p}\}$ and $\{\phi_{2,p}\}$ from $\{\phi_{1,l}^{\star}\}$ and $\{\phi_{2,l}^{\star}\}$ respectively with $P=d_\mathrm{max}$ and $p=1,2,\dots,P$
    \STATE Calculate $\{T_{\mathrm{c},t}\}$ using \eqref{29} and solve beamforming optimization problem \eqref{eq:optimization} to obtain $\{T_{\mathrm{g},t}\}$ based on $\{\psi_{1,p}\}$ and $\{\phi_{2,p}\}$
    \STATE Obtain final aRIS coefficients $\mathbf{\Phi}=\mathrm{diag}\{\mathbf{T}\otimes\mathbf{1}_{N_{a}}\}$
    \RETURN $\mathbf{p}_{\star},\mathbf{\Phi}$
\end{algorithmic}
\end{algorithm}

\begin{figure}[htbp]
\centerline{\includegraphics[width=0.98\textwidth]{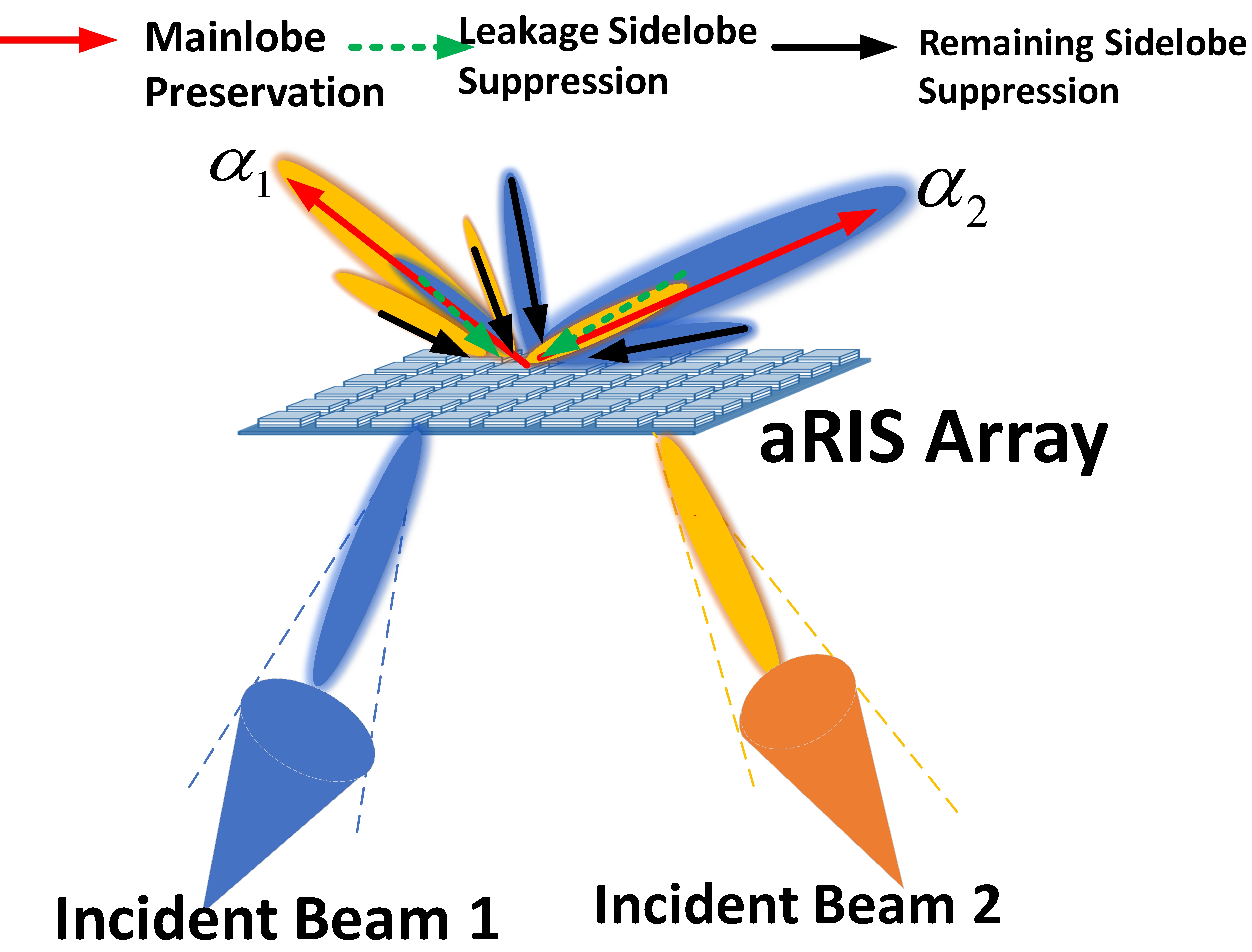}}
\vspace{-3mm} 
\caption{Independent multi-DoF beamforming.}
\label{fig:wet_end}
\vspace{-4mm} 
\end{figure}
After that, the design of $T_{\mathrm{g},t}$ becomes a multi-directional beamforming problem, that is, $P$ independent beams should be generated toward the receiver to support $P$ DoFs. A key challenge for this problem is \emph{inter‑beam sidelobe coupling}, which means the sidelobe of the beam in one direction may overlap the main lobe of another, causing leakage and corrupting the intended signal paths. To cope with this challenge, we propose to design $T_{\mathrm{g},t}$ by solving an optimization problem. In specific, defining the $P$ AoDs of resolvable paths between the aRIS and the receiver as $\psi_{2,p}$, the complex baseband pressure observed in any departure direction $\psi$ is $\eta(\psi) = \sum_{p}\eta_{p}(\psi)=\sum_{p}\mathbf{e}_{a}\left(\psi\right)^{H}\mathbf{T}^{(p)}_{\mathrm{g}}$, where $\mathbf{T}^{(p)}_{\mathrm{g}}=[T_{\mathrm{g},1}^{(p)},T_{\mathrm{g},2}^{(p)},\cdots,T_{\mathrm{g},N_{a}}^{(p)}]^{T}$ is the beamforming coefficient to generate beams toward $\psi_{2,p}$ which satisfies $T_{\mathrm{g},t}=\sum_{p=1}^{P}T_{\mathrm{g},t}^{(p)}$, and $\eta_{p}(\psi)$ is the complex baseband pressure achieved by $\mathbf{T}^{(p)}_{\mathrm{g}}$. Therefore, we formulate the following optimization problem to obtain $T_{\mathrm{g},t}$ as
\begin{subequations}\label{eq:optimization}
\begin{align}
    &\{T_{\mathrm{g},t}\}=\arg\min\enspace \beta \\
    &\text{s.t.}\enspace\,\, \left| \eta_{p}(\psi_{2,p}) \right| \geq G_i^{req}, \\
     &\quad\quad\left| \eta_{k}(\psi_{2,p}) \right| \leq \epsilon_{cross}, \quad k \neq p, \\
     &\quad\quad\left| \eta_{p}(\psi) \right| \leq \beta, \quad \forall \psi \in \left[-90^\circ, 90^\circ\right] \setminus \left\{\psi_{2,p}\right\}\\
     &\quad\quad\left| T_t \right| \leq G_{max}
\end{align}
\end{subequations}
where $G_i^{req}$ denotes the required main-lobe gain, $\epsilon_{cross}$ represents the allowable interference threshold, $G_{max}$ is the maximum transmission power constraint of the aRIS. The global constraint $\beta$ in (20d) further minimizes sidelobe levels, effectively reducing interference. Due to constraints involving magnitudes and inner products, the optimization in (\ref{eq:optimization}) forms a standard second-order cone programming (SOCP) problem, solvable by mature algorithms~\cite{domahidi2013ecos}. Consequently, this optimization ensures desired main-lobe directions while suppressing inter-beam leakage, thus achieving the upper bound of $\mathrm{DoF}_{\mathrm{RIS}}$ by fully utilizing the resolvable paths in \(R_{2}(\mathbf{p})\).

\begin{algorithm}[htbp]
\caption{Layered Ray-Tracing (LRT)}
\label{alg:joint}
\begin{algorithmic}[1]
    \STATE \textbf{Input:} Tx position $(r_t,z_t)$, Rx position $(r_r,z_r)$,
           SSP $c(z)$ discretised into $M$ layers $\{c_m,h_m\}_{m=1}^{M}$
    \STATE \textbf{Output:} The number of propagation paths between Tx and Rx $L$, the set of AoAs $\mathcal{A}_{a}=\{\phi_{l}\}$ and AoDs $\mathcal{A}_{d}=\{\psi_{l}\}$
    
    \STATE $L\leftarrow 0$, $\mathcal A_{a}\leftarrow\varnothing$, $\mathcal A_{d}\leftarrow\varnothing$
    \STATE \textbf{for} each departure angle $\alpha_0$ from Tx \textbf{do}
    \STATE \quad $m\leftarrow 1,\;\alpha\leftarrow\alpha_0,\;z\leftarrow z_{t}$
    \STATE \quad \textbf{while} $z<z_r$ \textbf{do}
    \STATE \quad\quad $\alpha\leftarrow\arccos\!\left(\frac{c_{m+1}\cos\alpha}{c_m}\right),z\leftarrow z+h_m,m\leftarrow m+1$
    \STATE \quad \textbf{end while}
    \STATE \quad Obtain AoA $\phi_{l}=\pi/2-\alpha$
    \STATE \quad $L\leftarrow L+1$, $\mathcal A_{a}\leftarrow\mathcal A_{a}\cup\{\phi_{l}\}$, $\mathcal A_{d}\leftarrow\mathcal A_{d}\cup\{\alpha_{0}\}$
    \STATE\textbf{end for}
    \RETURN $L$, $\mathcal{A}_{a}$,$\mathcal{A}_{d}$
            
\end{algorithmic}
\end{algorithm}

\vspace{-3mm}
\subsection{Joint aRIS Deployment and Beamforming Scheme}
\label{CC}

Based on the optimal deployment of aRIS and multi-DoF beamforming design which have been introduced, we propose a joint aRIS deployment and beamforming scheme. The pseudocode of the proposed scheme is shown in \textbf{Algorithm 1}. In the proposed scheme, the optimal Light-point $\mathbf{p}_{\star}$ is first selected as the deployment position of aRIS in step 3-13 according to the principle described in Section II. We adopt a \emph{layered ray‑tracing} (LRT) as shown in \textbf{Algorithm 2} to determine the number of paths, the AoAs, and the AoDs of the channel $\mathbf{H}_{1}$ and $\mathbf{H}_{2}$ in step 5 and 6. Specifically, the water column is discretised into $M$ homogeneous layers, within each of which the acoustic ray propagates straight. Snell’s law relates successive angles as $\frac{\cos\alpha_{m}}{c_{m}}
    =
    \frac{\cos\alpha_{m+1}}{c_{m+1}},\quad m=1,\dots,M\!-\!1$.
After the optimal Light-Point $\mathbf{p}_{\star}$ is found out, the coefficients of aRIS $\mathbf{\Phi}$ is obtained by utilizing the beamforming design introduced in Subsection IV-B in step 16-19. By deploying the aRIS at the optimal Light-Point and realizing multi-DoF beamforming, the proposed scheme can achieve efficient DoF-increasing and channel capacity enhancement.

\section{Dynamic Light-Point Tracking and Environmental Adaptation}\label{RISrobust}
Although we have already proposed an efficient joint aRIS deployment and beamforming scheme, it is for an static environment. To deal with the dynamic environment, in this section, we present the methodology for dynamic tracking the Light-Point and environmental adaptation using the integration of UUVs and aRIS, which can make the proposed scheme more robust in practical underwater scenarios.

\subsection{Light-Point Tracking Using Gaussian Ray Model}\label{AA}
In the proposed scheme, it would be ideal if the transmitter could transmit only along the propagation paths between the transmitter and the aRIS making the signals reach the aRIS. However, in practical systems, a transmitter cannot emit acoustic signals in a single direction. Instead, it can only emit a beam centered on a certain direction. Usually, the power distribution of this beam is considered to be Gaussian distribution \cite{zhaoyangMIMO}. In specific, sound intensity \((I(r)\) at distance from the center of the beam \(r\) is given by $I(r) = I_0e^{-\frac{r^2}{\sigma^2}}$, where \(I_0\) is the maximum intensity and \( \sigma \) represents the width of the beam. Due to dynamic oceanic conditions (e.g., turbulence and wave motion), the initial aRIS placement may deviate slightly from the optimal Light-Point, leading to suboptimal reception \cite{9053309}. If the aRIS offset from the beam center is \(\Delta r\), the received power is proportional to the intensity at distance \(\Delta r\), expressed as \cite{jensen2011computational}:$\label{eq:Ax}A_{\text{received}} = I(\Delta r)N_{a}^{2}\Delta_{a}^{2}\lambda_{c}^{2}$.
Since the aRIS can estimate the angles‑of‑arrival (AoAs)~\cite{wang2023designing} by means of its absorption units, it is able to infer the local energy gradient at $\mathbf x=[r,z]^{T}$  and accordingly adjust its pose.   
To obtain gradient \emph{online}, we adopt a two‑sample finite‑difference scheme:  
\vspace{-2mm}
\begin{subequations}\label{eq:grad_est}
\begin{align}
    A_0 &\triangleq I\!\bigl(r\bigr)N_{a}^{2}\Delta_{a}^{2}\lambda_{c}^{2},\\
    A_{i+} &\triangleq I\!\bigl(r+\delta e_i\bigr)N_{a}^{2}\Delta_{a}^{2}\lambda_{c}^{2},\qquad i\in\{r,z\},\\
    \hat{\mathbf g}_i &\triangleq \frac{A_{i+}-A_0}{\delta}\approx
        \frac{\partial A_{\text{received}}}{\partial r_i}.
\end{align}
\end{subequations}
where $\delta$ is a small probing displacement along the unit vector $e_i$.  
\textit{Note that whenever the probe step $+\delta e_i$ moves \emph{away} from the beam center, we have $A_{i+}<A_0$, so $\hat g_i<0$; conversely, if it moves \emph{toward} the center, $\hat g_i>0$.  Thus $\hat{\mathbf g}=[\hat g_r,\hat g_z]^{T}$ always points toward the direction of \textbf{steepest ascent} of $A_{\text{received}}$.}  
The aRIS position is then updated by $\mathbf x^{(k+1)}=\mathbf x^{(k)}+\eta\,\hat{\mathbf g}^{(k)}$, with $\eta$ the step size and $k$ the ping index.  

\subsection{UUV–aRIS Integration and Optimization}\label{BB}
\begin{figure}[htbp]
\centerline{\includegraphics[width=0.98\textwidth]{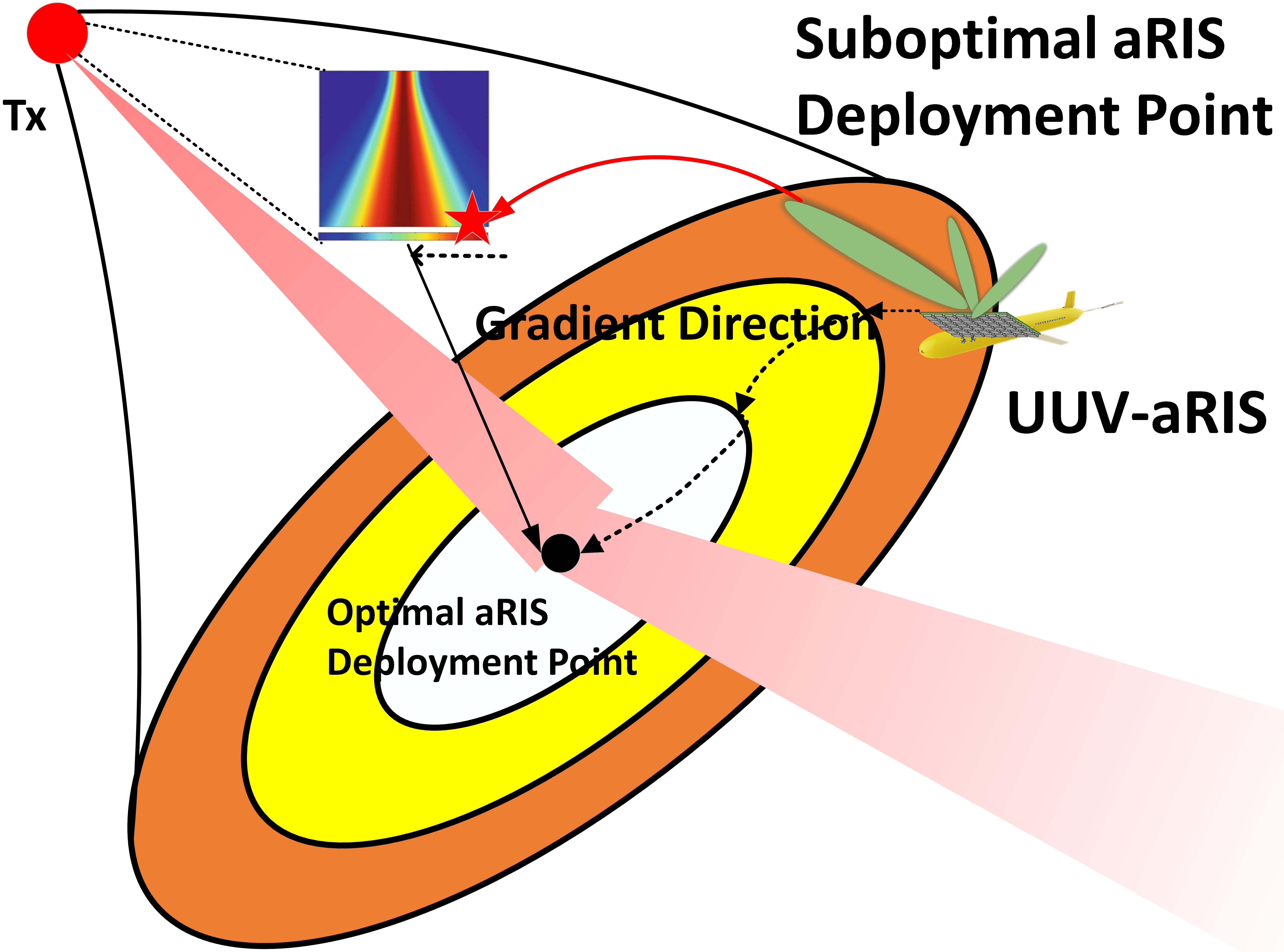}}
\vspace{-3mm} 
\caption{Dynamic beam-tracking process.}
\label{fig:shallow_sea_coverage} 
\vspace{-8mm}
\end{figure}

As shown in Fig.~7, integrating the aRIS with the UUV enables real‑time position adaptation without adding communication overhead~\cite{10.1145/3491315.3491324}.  
The UUV first brings the aRIS to a neighbourhood of the Light‑Point; thereafter, the fine adjustment is accomplished by the gradient iteration.  
Below we establish the convergence of this procedure.
The receive‑power surface \(A(\mathbf x)\) in~\eqref{eq:Ax} is a \emph{single‑peaked} (log‑concave) Gaussian whose gradient is globally Lipschitz with constant \(L=2I_0/\sigma^{2}\).  
If a constant step size satisfying \(0<\eta<2/L\) is used in the update rule~(34), each iteration strictly increases \(A(\mathbf x)\) yet keeps it below the finite ceiling \(A(\mathbf p_{\star})\); the sequence \(\{A(\mathbf x_k)\}\) therefore climbs monotonically and eventually saturates.  
Because the only point where the gradient vanishes is the Light‑Point \(\mathbf p_{\star}\), the offsets \(\mathbf x_k\) necessarily settle there, i.e.\ \(\mathbf x_k\!\to\!\mathbf p_{\star}\).  
This gradient-based adaptation efficiently drives the aRIS towards the optimal deployment point, maximizing the signal reception and ensuring robust communication in dynamic marine environments\cite{yuan2016convergence}.

\section{Validation And Evaluations}\label{Sim}
In this section, we use Bellhop-based simulations to validate the performance of the aRIS-aided UWA MIMO system utilizing the designed ASTAR aRIS and the proposed joint aRIS deployment and beamforming scheme. An aRIS-aided UWA MIMO system is considered in deep-sea and shallow-sea scenarios, using typical SSPs, with $N_{t}=4$, $N_{r}=4$, center frequency \( f_c = 9\,\text{kHz} \).

\subsection{Validation on the Deployment Principle of aRIS}\label{AA}

Fig.~8 illustrates simulation results for two scenarios. In the deep-sea scenario, deploying aRIS at the intersection as Light-Point redirects two distinct acoustic paths (blue and green curves), clearly separating their AoA at the receiver and effectively increasing spatial DoFs by two. For the shallow-sea case, the aRIS strategically redirects intersecting LOS, surface-reflected, and seabed-reflected rays (black, purple, and red curves), creating three distinguishable propagation channels and enhancing DoFs by three. These simulations visually confirm that our proposed deployment principle improves path distinguishability, effectively enhancing the spatial DoF.

\begin{figure}[htbp]
\centerline{\includegraphics[width=0.98\textwidth]{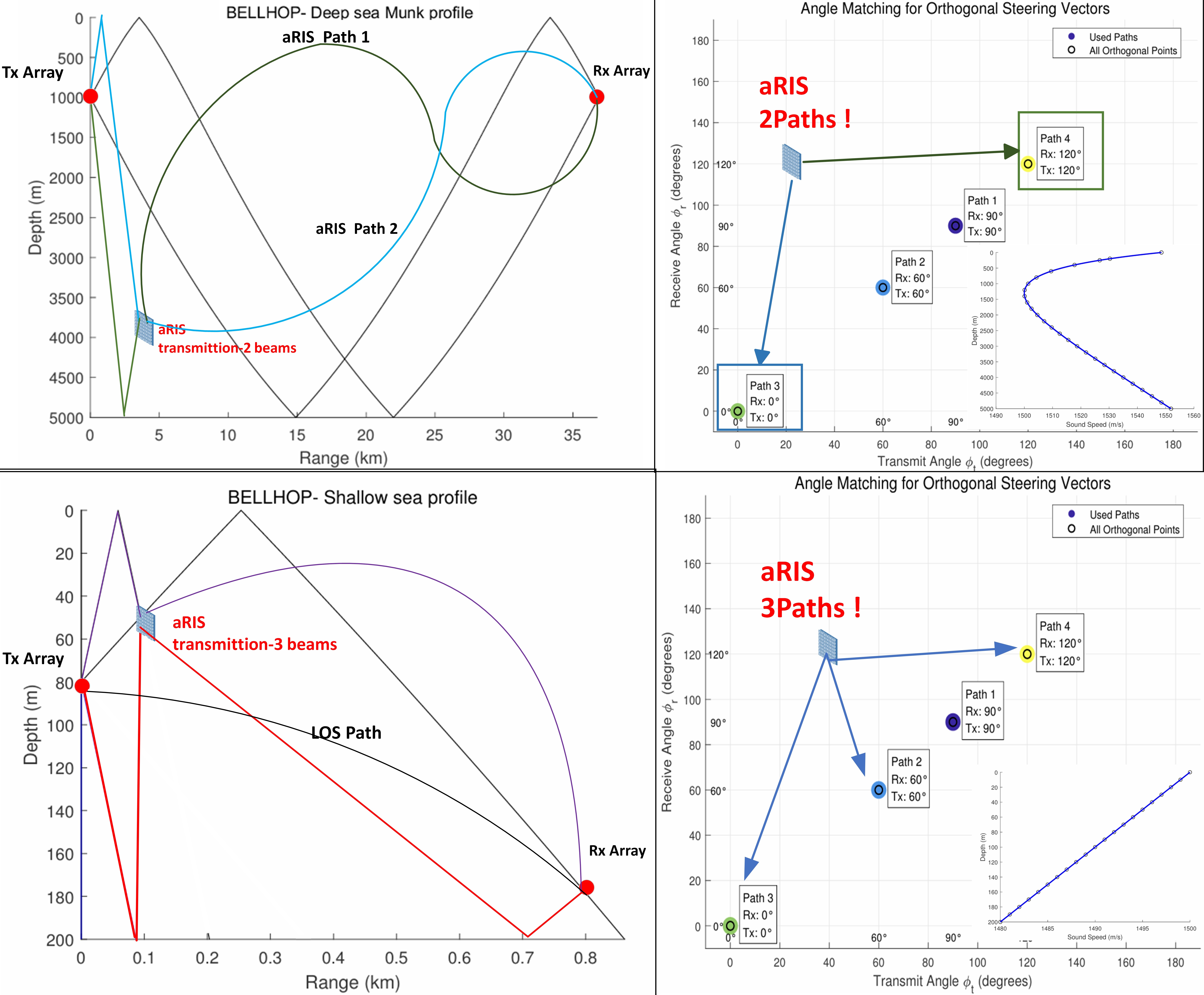}}
\caption{Deployment of aRIS, path visualization and AoA distinguishability.}
\label{fig:deepsea_coverage}
\vspace{-4mm} 
\end{figure}

\subsection{Performance of Independent Multi-Dof Beamforming}\label{BB}
To validate the effectiveness of our proposed independent multi-DoFs beamforming approach, we consider a 64-element ($N_a=8 $) planar aRIS. Two desired transmission beam directions were set to $\theta_1 = -30^\circ$ and $\theta_2 = 30^\circ$, respectively. The required main-lobe gains ($G_i^{req}$) is set to $0.8$, with cross-interference suppression ($\epsilon_{cross}$) set as $0.01$, and the maximum allowable transmission coefficient magnitude ($G_{max}$) limited to $1$.
In Fig.~9 (left), we present a comparison between the optimized multi-beam pattern obtained through our SOCP-based optimization method and the conventional beamforming scheme, which directly sums transmission weights without explicit interference suppression. Clearly, our optimized beamforming achieves main-lobe gains close to the desired directions, with sidelobe levels considerably lower than those of the conventional method.  Fig.~9  (right) further provides a two-dimensional visualization of the optimized beam pattern. This figure clearly illustrates two distinct main lobes oriented precisely toward the desired directions ($\theta_1 = -30^\circ$ and $\theta_2 = 30^\circ$), with minimal sidelobe leakage.

\begin{figure}[htbp]
\centerline{\includegraphics[width=0.98\textwidth]{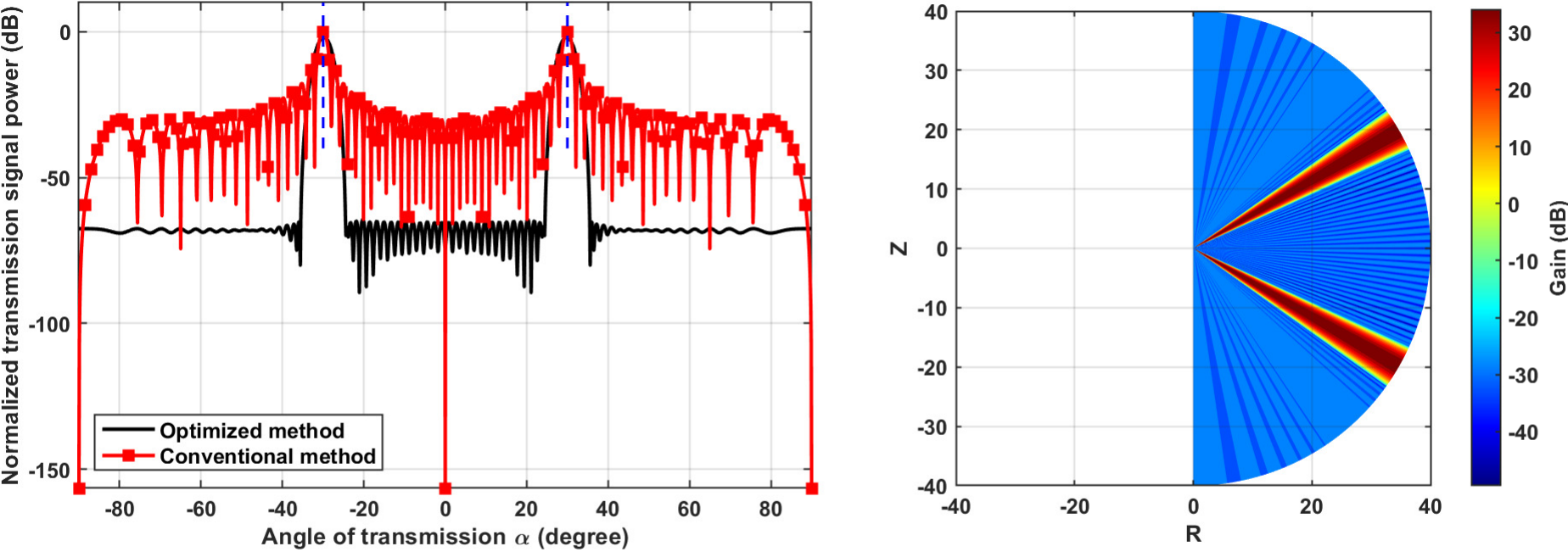}}
\vspace{-3mm} 
\caption{aRIS multi-beam transmission pattern.}
\label{fig:aRIS multi-beam transmission pattern.}
\vspace{-5mm} 
\end{figure}

\subsection{Dynamic Light-Point Tracking}\label{CC}
We validate the proposed adaptive Light-Point scheme through numerical simulations using a 2D Gaussian acoustic energy field, with maximum intensity at the beam center $(0,0)$, standard deviation $\sigma=50\,\text{m}$, and a deployment radius $R_{\text{max}}=100\,\text{m}$. The aRIS is initially placed randomly at distances of approximately $60$–$70\,\text{m}$ from the center. Using gradient ascent (step size $\eta=4.0$, max iterations $=50$, tolerance $=10^{-4}$), we iteratively adjust the aRIS position toward the optimal Light-Point.
Fig.~10 illustrates the optimization process. Initially, random placement yields only 30–40\% of the maximum achievable energy. The adaptive scheme repidly and robustly converges, moving the aRIS closer to the optimal location (Fig.~10, left), ultimately capturing over 95\% of the maximum energy (Fig.~10, right).

\begin{figure}[htbp]
\centerline{\includegraphics[width=0.98\textwidth]{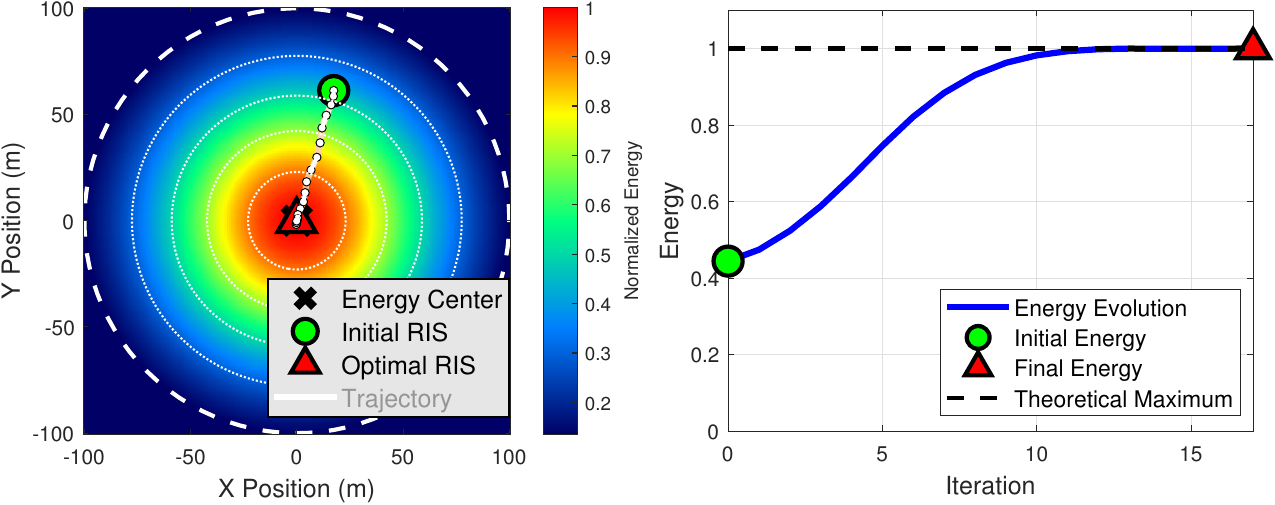}}
\caption{Dynamic Beam-Tracking and Environmental Adaptation via UUV-aRIS Integration.}
\label{fig:deepsea_coverage}
\vspace{-2mm} 
\end{figure}

\vspace{-1.5mm}
\subsection{Channel Capacity of ARIS-aided UWA MIMO Systems}\label{DD}
We further evaluate the channel capacity enhancement of the proposed joint aRIS deployment and beamforming scheme using the designed ASTAR aRIS in both shallow-sea and deep-sea environments. The MIMO channel capacity is computed as $C=\log_2\det\!\left(\mathbf{I}_{N_r}+\frac{\rho\,\gamma_{\text{track}}(k)}{N_t}\mathbf{H}_{\text{eff}}\mathbf{H}_{\text{eff}}^{H}\right)$, where \(\rho\triangleq P_t/\sigma_n^2\) denotes the nominal transmit SNR with total transmit power \(P_t\) and per-element noise variance \(\sigma_n^2\), \(N_t=N_r=4\), and \(\mathbf{H}_{\text{eff}}=\mathbf{H}+\mathbf{H}_2\mathbf{\Phi}\mathbf{H}_1\) represents the effective channel accounting for aRIS effects. The factor \(\gamma_{\text{track}}(k)\in(0,1]\) captures the received-energy gain enabled by dynamic Light-Point tracking and is defined as \(\gamma_{\text{track}}(k)\triangleq A_{\text{received}}(k)/A_{\max}\), where \(A_{\text{received}}(k)\) is the energy collected after \(k\) tracking iterations and \(A_{\max}\) is the corresponding theoretical maximum attained at the Light-Point.
Fig.~11 (left) depicts the shallow-sea results, where integrating aRIS yields substantial capacity gains over the baseline without aRIS. In particular, when the aRIS is optimally deployed at the identified Light-Point and Light-Point tracking is effectively performed, the channel capacity increases by approximately \(265\%\) relative to the baseline. Our evaluations also highlight the importance of accurate tracking: suboptimal aRIS positioning can considerably reduce the achievable capacity. Fig.~11 (right) shows similar trends in the deep-sea scenario, where the proposed optimal deployment and beamforming achieve about \(170\%\) capacity enhancement over the baseline.

\begin{figure}[htbp]
\centerline{\includegraphics[width=0.98\textwidth]{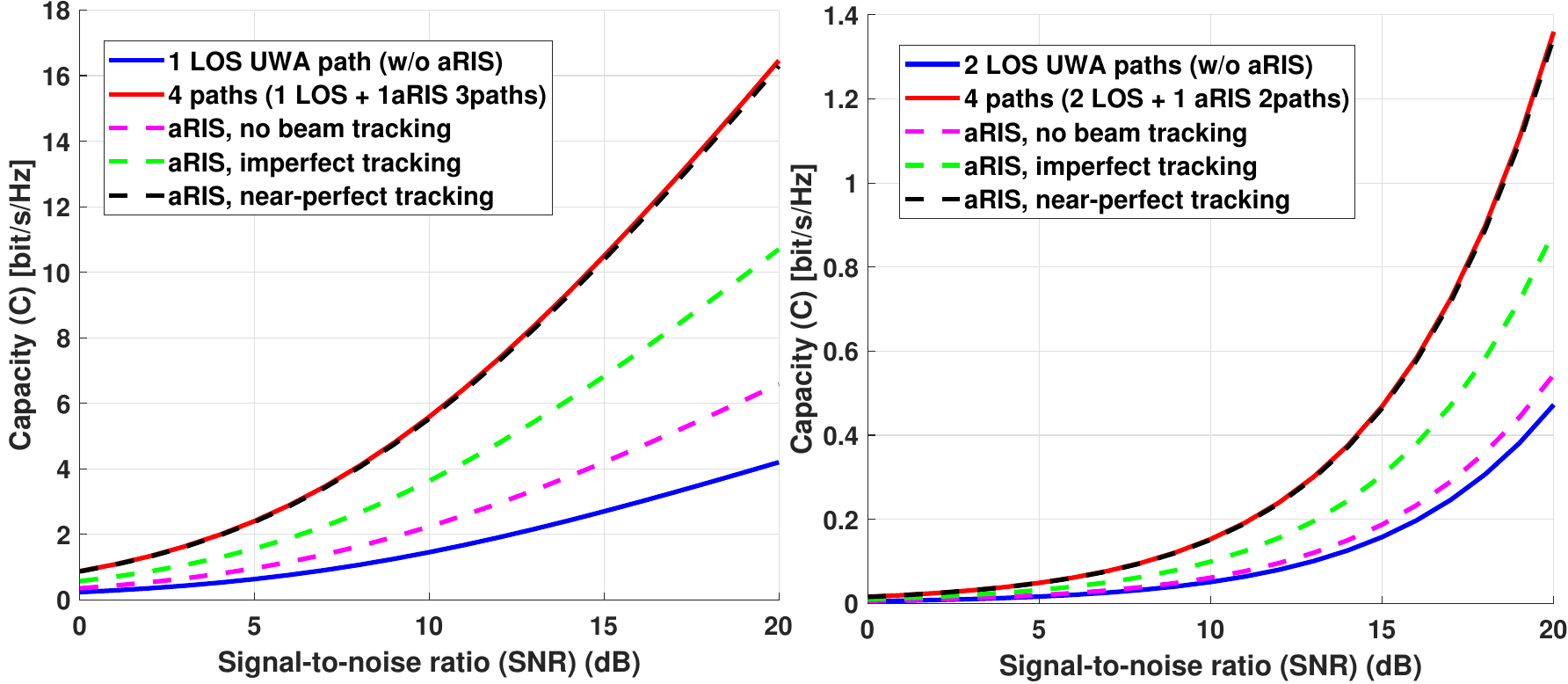}}
\vspace{-3mm} 
\caption{The capacity of aRIS aided 4$\times$4 MIMO in  shallow-sea scenarios(left); The capacity of aRIS aided 4$\times$4 MIMO in deep-sea scenarios(right).}
\label{fig:shallowseacase}
\vspace{-8.3mm} 
\end{figure}

\vspace{-2.5mm}
\section{Conclusions}\label{Con}

This paper presents an effective approach to enhancing underwater acoustic (UWA) MIMO channel capacity and spatial degrees-of-freedom (DoFs) through the innovative use of acoustic Reconfigurable Intelligent Surfaces (aRIS). 
A multipath-based DoF model and the \textit{Light-Point principle} are introduced, optimizing aRIS deployment to maximize spatial DoFs by leveraging acoustic ray intersections.
An active simultaneous transmitting and reflecting (ASTAR) aRIS structure is proposed, introducing two and three additional DoFs in deep-sea and shallow-sea scenarios, respectively. Additionally, an adaptive beam-tracking scheme integrating unmanned underwater vehicles (UUVs) ensures robustness under dynamic oceanic conditions.
Extensive simulations validate the framework, demonstrating substantial channel capacity improvements—approximately \textbf{265\%} in shallow-sea and \textbf{170\%} in deep-sea scenarios.

\section*{Acknowledgement}

The work of Longfei Zhao and Zhi Sun are supported by the National Natural Science Foundation of China, with a Grant of No. 62271284 for the project "Towards Acoustic Reconfigurable Intelligent Surface for High-data-rate Long-range Underwater Communications".
    
The work of Jingbo Tan and Jintao Wang are supported by the National Natural Science Foundation of China, with a Grant of No. 62401315 for the project ”Acoustic Reconfigurable Intelligent Surface Based High-speed Mid-Long-Range Deep-Sea Underwater Acoustic Communication Theory and Key Technologies”.

The work of Ian F. Akyildiz is supported by the Icelandic Research Fund administered by Rannís – the Icelandic Centre for Research, with a Grant of Excellence No. 239994-051 for the project “HAF: Under-water Robotics Sensor Networks with Multi-Mode Devices and Remote Power Charging Capabilities.

\newpage
\bibliographystyle{IEEEtran}
\bibliography{reference}

\end{document}